\documentclass[sigconf,natbib=true]{acmart}
\AtBeginDocument{%
  }

\setcopyright{acmlicensed}
\usepackage{enumitem}
\usepackage{booktabs}
\usepackage{colortbl}
\usepackage{graphicx}

\usepackage{amsthm}

\newtheorem{proposition}{Proposition}
\usepackage{mathtools}

\newtheorem{assumption}{Assumption}

\usepackage{multirow}
\usepackage{amsmath}
\usepackage{amsthm}

\newtheorem{definition}{Definition}

\usepackage{colortbl}
\definecolor{lightgray}{gray}{0.9}
\usepackage{subfigure}
\copyrightyear{2018}
\acmYear{2018}
\acmDOI{XXXXXXX.XXXXXXX}
\acmConference[Conference acronym 'XX]{Make sure to enter the correct
  conference title from your rights confirmation email}{June 03--05,
  2018}{Woodstock, NY}
\acmISBN{978-1-4503-XXXX-X/2018/06}

\begin{document}

\title{When Recommendation Denoising Meets Popularity Bias: Understanding and Mitigating Their Interaction}

\author{Guohang Zeng}
\affiliation{%
  \institution{Australian Artificial Intelligence Institute}
  \institution{University of Technology Sydney}
  \city{Sydney}
  \country{Australia}}
\email{guohang.zeng@student.uts.edu.au}

\author{Jie Lu}
\authornote{Corresponding Author}
\affiliation{%
\institution{Australian Artificial Intelligence Institute}
  \institution{University of Technology Sydney}
  \city{Sydney}
  \country{Australia}}
\email{jie.lu@uts.edu.au}

\author{Guangquan Zhang}
\affiliation{%
\institution{Australian Artificial Intelligence Institute}
  \institution{University of Technology Sydney}
  \city{Sydney}
  \country{Australia}}
\email{guangquan.zhang@uts.edu.au}


\begin{abstract}
Implicit feedback is the dominant data source for recommender systems, but behavioral logs are often contaminated by false-positive interactions caused by mis-clicks, biased exposure, and interface effects. Denoising recommendation methods improve robustness by down-weighting or filtering interactions suspected to be noisy, often relying on the small-loss heuristic. We revisit this heuristic through the lens of popularity bias. Tail-item positives can be harder to fit because they are sparsely observed, and thus may receive larger losses even when they reflect genuine user preference. Under such popularity-dependent loss patterns, monotone loss-based reweighting can suppress clean-but-hard tail signals and increase the head-tail imbalance in effective supervision.

We formalize this interaction through the effective head-tail signal ratio induced by denoising weights and derive a conditional reallocation result: when the loss distribution of tail positives is right-shifted relative to that of head positives, small-loss reweighting increases the effective head-tail signal ratio compared with ERM. Motivated by this analysis, we propose Popularity-Aware Denoising (PAD), a lightweight plug-in framework that modulates denoising strength by item popularity. PAD applies stronger denoising to highly exposed items while being more conservative on tail items, preserving more clean-but-hard long-tail signals. Experiments on three datasets and three backbones show that PAD generally improves over representative denoising baselines and provides favorable accuracy-diversity tradeoffs, especially on MF-style recommenders. 
\end{abstract}





\maketitle

\section{Introduction}

Modern recommender systems are predominantly trained from \emph{implicit feedback} \cite{hu2008collaborative, joachims2017accurately, rendle2012bpr}--behavioral logs such as clicks, purchases, and reviews--rather than explicit ratings. Compared with explicit feedback, implicit feedback is much easier to collect in real-world applications and therefore serves as the primary supervision source for recommendation models. However, such logs only indirectly reflect user preference and are affected by multiple sources of noise, including false-positive interactions caused by mis-clicks \cite{tolomei2019you}, biased exposure \cite{baeza2020bias, chen2023bias}, and other presentation effects. To cope with these noisy positives, \emph{denoising recommendation} \cite{wang2021denoising, wang2023efficient, gao2022self, chua2024unified, zhang2025personalized, zeng2025we} has emerged as an important line of work, aiming to reduce the impact of unreliable interactions during training.

From a statistical perspective, denoising recommendation can be viewed as estimating the posterior reliability of an observed positive interaction, i.e., $p(Y^*_{u,i}=1 \mid \bar{Y}_{u,i}=1,u,i)$, where $Y^*$ denotes the unobserved clean preference label and $\bar{Y}$ denotes the observed implicit interaction. Since $Y^*$ is unavailable in ordinary implicit-feedback logs, this posterior is not identifiable from observations alone. As a result, practical denoising methods inevitably rely on heuristic assumptions to approximate which observed positives are likely to be clean.

A widely used principle in denoising recommendation and label-noise learning is the \emph{small-loss criterion} \cite{han2018co, yu2019does}: interactions with smaller training losses are treated as more reliable, while large-loss positives are down-weighted or removed. This heuristic is useful, but it also has an important blind spot. A large loss does not necessarily indicate a false-positive interaction; it may also correspond to a clean but hard positive signal that the model has not yet learned. This distinction is especially important in recommendation, where item popularity strongly shapes both data availability and learning difficulty.

In particular, tail-item positives are often supported by fewer observations than head-item positives. Their representations are therefore more difficult to estimate, and clean preferences for tail items may incur larger losses during training. This creates a potential interaction between denoising and popularity bias: if a loss-based denoiser treats high-loss positives as unreliable, it may suppress not only noisy positives but also clean-but-hard long-tail signals. The resulting effective supervision can shift toward head items, worsening the head--tail imbalance that already exists in implicit-feedback recommendation.

\begin{figure}[t]
  \centering
  \includegraphics[width=0.47\textwidth]{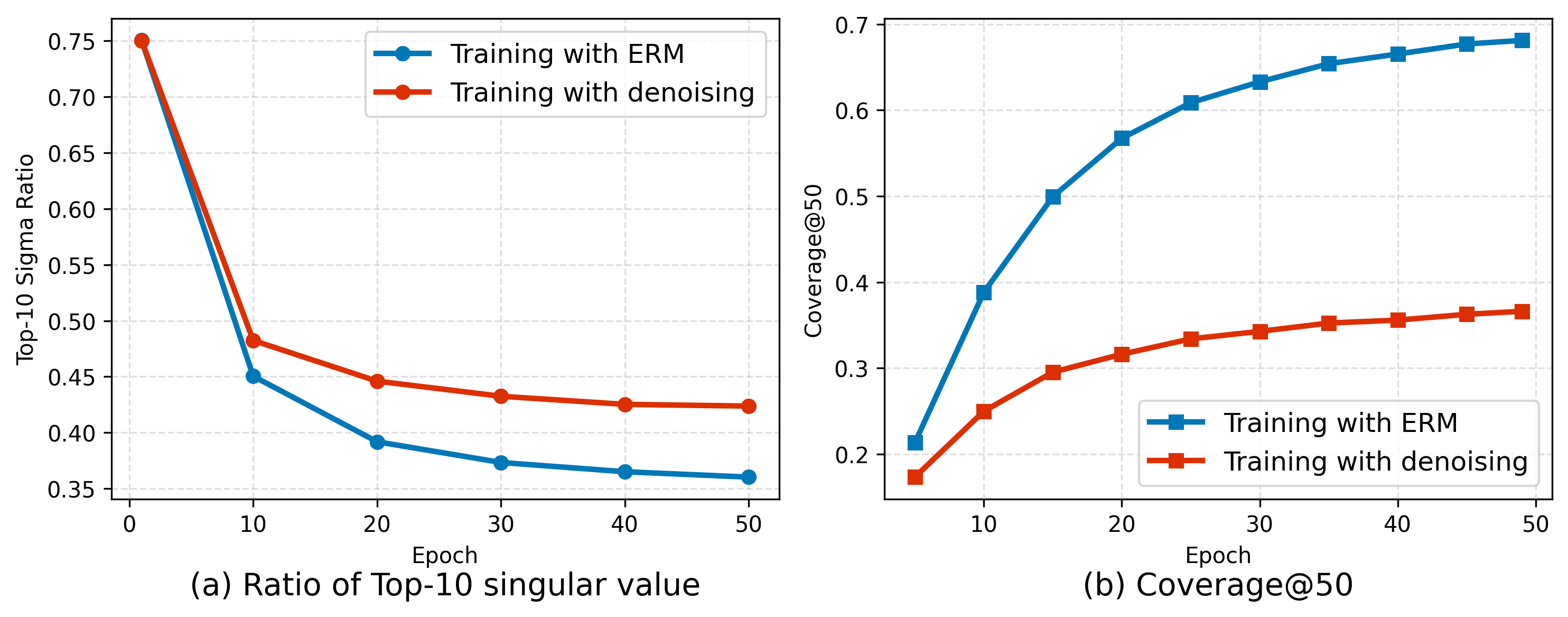}
  \caption{A motivating example of the interaction between loss-based denoising and popularity bias. On GMF trained on Yelp, denoising changes both representation concentration and recommendation coverage compared with ERM. This example motivates our conditional analysis rather than serving as universal evidence. (a) Spectral concentration of item representations, measured by the Top-10 singular-value mass ratio $\sum_{j=1}^{10}\sigma_j / \sum_j \sigma_j$. (b) Coverage@50.}
\label{fig:fig1}
\vspace{-2em}
\end{figure}

This raises the central question of this paper: \emph{when loss-based denoising is applied to implicit-feedback recommendation, how does it interact with popularity-induced head--tail imbalance?} We answer this question through a conditional analysis. Rather than claiming that denoising universally worsens popularity bias, we characterize a concrete failure mode: when tail positives have a right-shifted loss distribution relative to head positives, monotone small-loss reweighting increases the effective head--tail signal ratio compared with ERM. Figure~\ref{fig:fig1} provides a motivating example. In one representative setting, a loss-based denoising method changes both representation concentration and recommendation coverage compared with ERM, suggesting that denoising can reshape popularity-related behavior.

Motivated by this analysis, we propose \emph{Popularity-Aware Denoising} (PAD), a lightweight plug-in framework that modulates denoising strength by item popularity. PAD keeps stronger denoising on highly exposed head items, where false-positive interactions are more likely to arise from exposure and presentation effects, while applying more conservative denoising to tail items to avoid suppressing sparse but informative positives. Importantly, PAD is not intended to replace existing denoising estimators; instead, it wraps a base denoising weight with a popularity-conditioned gate.

Empirically, we evaluate PAD on three datasets, three backbone models, representative denoising baselines, and both BCE and BPR objectives. The results show that PAD generally improves over uniform loss-based denoising baselines and often provides a better accuracy--diversity tradeoff, especially on MF-style backbones such as GMF and NeuMF. At the same time, LightGCN+ERM remains a strong reference in several settings. We treat this not as a contradiction, but as an important boundary case: graph propagation may already smooth sparse signals and reduce the need for explicit denoising. This boundary is consistent with our conditional formulation.

In summary, our contributions are as follows:
\begin{itemize}[leftmargin=*]
    \item \textbf{Interaction perspective:} We identify recommendation denoising and popularity bias as interacting mechanisms rather than isolated issues, and study how loss-based denoising can alter the allocation of effective positive supervision between head and tail items.
    \item \textbf{Conditional characterization:} We provide a formal characterization showing that, when tail positives have a right-shifted loss distribution, monotone small-loss reweighting increases the effective head--tail signal ratio relative to ERM.
    \item \textbf{Popularity-aware mitigation:} We propose PAD, a lightweight plug-in gating framework that modulates denoising strength by item popularity to preserve clean-but-hard tail signals while retaining denoising on highly exposed items.
    \item \textbf{Boundary-aware empirical study:} Using existing experiments across datasets, backbones, denoising baselines, and a BPR compatibility check, we show that PAD often improves over denoising baselines while also revealing boundary cases where ERM remains highly competitive.
\end{itemize}

\section{Preliminaries}



We consider the implicit-feedback setting in recommender systems. Let $\mathcal{U}=\left\{u_1, u_2, \ldots \right\}$ denote the set of users and $\mathcal{I}=\left\{i_1, i_2, \ldots \right\}$ represent the set of items, respectively. The supervision information is derived from an interaction matrix $\mathbf{Y} \in \{0, 1\}^{|\mathcal{U}| \times |\mathcal{I}|}$, where $y_{ui}=1$ indicates that user $u$ has interacted with item $i$, and $y_{ui}=0$ indicates otherwise. Since recommendation with implicit feedback corresponds to certain ranking and information retrieval problems, we are concerned with the following ranking metrics \cite{yang2018unbiased}.
\begin{equation}
\mathcal{R}_{ranking}=\frac{1}{|\mathcal{U}|} \sum_{u \in \mathcal{U}} \sum_{i \in \mathcal{I}} P\left(Y_{u, i}=1\right) \cdot c\left(\widehat{Z}_{u, i}\right),
\label{eq:rel}
\end{equation}
where $\widehat{Z}_{u,i}$ denote the predicted ranking of item $i$ for user $u$, and the function $c( \cdot )$ denotes the per-sample contribution induced by a ranking metric (e.g., top-$K$ hit, NDCG gain, etc). Since ranking metrics are often non-smooth, many of them contain non-differentiable/combinatorial structures such as $\mathbb I[\cdot]$ and $\mathrm{rank}(\cdot)$, making direct gradient-based optimization on $\widehat{Z}$ difficult. Hence, training typically relies on differentiable surrogate losses \cite{bruch2019analysis,di2025theoretical} such as Binary Cross-Entropy (BCE) loss and Bayesian personalized ranking (BPR) loss. We present the main derivation with the commonly used point-wise BCE loss, since many loss-based denoising methods for implicit-feedback recommendation instantiate their reweighting rules in this setting. We later include a BPR compatibility check to examine whether the proposed gating idea can also be applied to a pair-wise objective. Define the per-sample BCE loss as
$\delta(y,\hat p)= - (Y\log \hat p + (1-Y)\log(1-\hat p) )$, 
which can be equivalently decomposed into two terms:
$\delta^{(1)}_{u,i}=-\log \hat p_{u,i}$ and $\delta^{(0)}_{u,i}=-\log(1-\hat p_{u,i})$, and $\hat p_{u,i} \in(0,1)$ is the output of the model. Let the ``clean label'' be $Y^*_{u,i}\in\{0,1\}$. The empirical risk is
\begin{equation}
\mathcal L_{BCE}({D^*})
=\frac{1}{|\mathcal D^*|}\sum_{(u,i)\in\mathcal D^*}
\Big[
Y^*_{u,i}\,\delta^{(1)}_{u,i}
+\big(1-Y^*_{u,i}\big)\,\delta^{(0)}_{u,i}
\Big],
\label{eq:BCE_loss}
\end{equation}
where $\mathcal D^*$ denotes the training set with clean interactions $Y^*$. Minimizing BCE corresponds to maximum likelihood estimation of the posterior $p^*_{u,i}=P(Y^*_{u,i}=1 \mid u,i )$. For ranking rewards of the form in Eq.~\eqref{eq:rel}, ranking items by $p^*_{u,i}$ is Bayes-optimal \cite{schutze2008introduction, cossock2008statistical}. Hence, minimizing Eq. (\ref{eq:BCE_loss}) provides a surrogate for maximizing Eq. (\ref{eq:rel}).



\subsection{Denoising Recommendation}
In the denoising recommendation setting, we consider that the observed data $(u, i, \bar{Y})$ may not correspond to the true interaction $(u, i, Y^*)$. We denote the observed interactions set as $\mathcal{\bar{D}}=\left\{\left(u, i, \bar{Y}_{ui}\right) \mid u \in \mathcal{U}, i \in \mathcal{I}, \bar{Y}_{ui} \in \{0,1\} \right\}$. Note that due to the existence of noisy interactions, there is an inconsistency between $D^*$ and $\bar{D}$. Specifically, we consider that $\bar{D}$ contains noisy interactions, which can be represented as $\left\{(u, i) \mid Y^*=0 \wedge \bar{Y}=1\right\}$. These noisy interactions are typically introduced by users' accidental clicks or position bias. Given that the observed interaction data are not equal to the true interaction data, \cite{wang2021denoising} provided a formal definition of \textit{denoising recommendation training task} as:
\begin{equation}\small
\begin{aligned}
& {\Theta}^{\ast} = \min \mathcal{L}_{BCE}\big(denoise(\mathcal{\bar{D}})\big),
\end{aligned}
\label{eq:definision_denoise}
\end{equation}
where $denoise(\bar{\mathcal{D}})$ denotes applying a denoising method to the observed implicit feedback (e.g., pruning noisy interactions) during training, so as to learn reliable parameters $\Theta^{\ast}$. Conceptually, denoising aims to learn users' true preferences from noisy data.


\section{A Conditional Analysis of Denoising--Popularity Interaction}

In this section, we analyze when the inductive bias of loss-based denoising can introduce an additional head--tail imbalance in effective supervision. Our goal is not to claim that every denoising method always worsens popularity bias. Instead, we characterize a sufficient regime under which monotone small-loss reweighting reallocates positive training signal from tail items toward head items.

\subsection{Loss-Based Denoising as Heuristic Reweighting}
We first provide a formal view of denoising recommendation. Let the observed interaction be denoted by $\bar{Y}$, which may differ from the clean label $Y^*$. In this work, we focus on false-positive samples, i.e., observed positives that do not correspond to clean preferences. Although false negatives are also common in implicit-feedback recommendation, they are usually handled by negative sampling or negative mining and are beyond the scope of denoising positive feedback.

A broad class of denoising recommendation methods can be written as weighted empirical risk minimization:
\begin{equation}
\mathcal L_{\mathrm{denoise}}(\bar{D})
=\frac{1}{|\mathcal {\bar{D}}|}\sum_{(u,i)\in\mathcal{\bar{D}}}
\Big[
 w(u,i)\,\bar Y_{u,i}\,\delta^{(1)}_{u,i}
+\big(1-\bar Y_{u,i}\big)\,\delta^{(0)}_{u,i}
\Big],
\label{eq:loss_denoise}
\end{equation}
where $w(u,i)$ is a correction weight for an observed positive pair. Ideally, $w(u,i)$ should approximate the reliability posterior
\begin{equation}
 w(u,i) \approx P(Y^*_{u,i}=1 \mid \bar Y_{u,i}=1,u,i),
\label{eq:ideal_weight}
\end{equation}
so that the expected risk over noisy observations better matches the clean risk:
\begin{equation}
\mathbb{E}_{(u,i,\bar{Y})}\!\left[ \mathcal L_{\mathrm{denoise}}(\mathcal {\bar{D}}) \right]
\approx
\mathbb{E}_{(u,i,Y^{*})}\!\left[ \mathcal L_{BCE}(\mathcal{D}^*) \right].
\label{eq:L_denoise_eq_L_ERM}
\end{equation}
However, since $Y^*_{u,i}$ is unobserved, Eq.~\eqref{eq:ideal_weight} cannot be directly estimated from implicit logs. Practical denoising methods therefore rely on heuristic proxies for reliability.

\begin{assumption}[Small-loss criterion \cite{han2018co}]
For a positive interaction $(u,i)\in\mathcal{D}^+$, the positive loss $l^{+}(u,i)$ is monotonically related to the posterior probability that $(u,i)$ is a clean positive sample; equivalently, $\mathbb{P}((u,i)\ \text{is clean}\mid l^{+}(u,i))$ decreases as $l^{+}(u,i)$ increases.
\end{assumption}

This assumption should be understood as a modeling heuristic rather than a property of the data-generating process. It is nevertheless the design principle behind many denoising methods: high-loss positives are treated as less reliable and are assigned lower weights or removed. Such a heuristic can improve robustness against false positives, but it can also penalize genuinely positive interactions that are hard to fit.

\subsection{Popularity-Dependent Loss Patterns}
\begin{figure}[t]
  \centering
  \includegraphics[width=0.46\textwidth]{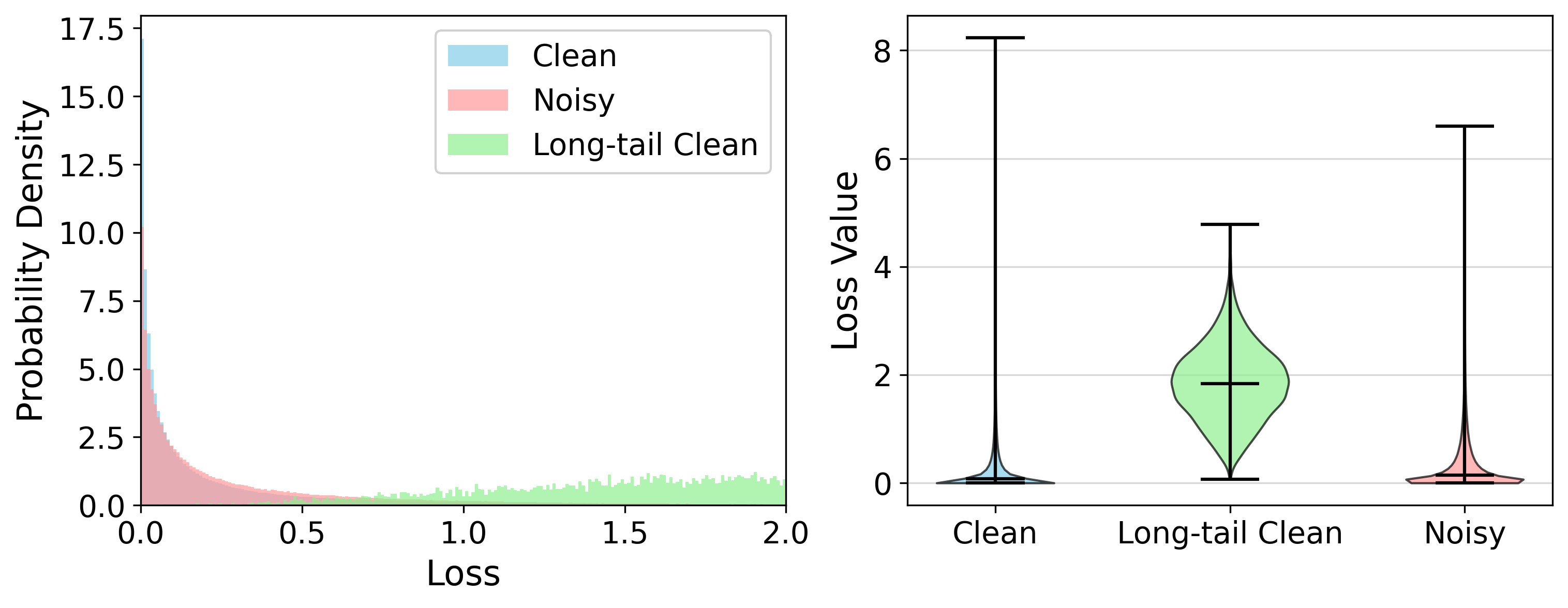}
  \caption{An illustrative loss diagnostic in one representative setting. We compare the loss distributions of clean positives, clean tail positives, and noisy positives for GMF on Yelp at an early training epoch. The figure illustrates that clean tail positives can overlap with high-loss/noisy regions, motivating our conditional analysis of popularity-dependent loss patterns. It is not intended as universal evidence across all datasets, backbones, or training epochs.}
  \label{fig:loss_ditribution}
\end{figure}

Before presenting the formal result, we provide a diagnostic example to illustrate why loss alone can be an imperfect proxy for noise in sparse tail regions. We train a GMF model on Yelp with ERM and visualize the loss distributions of clean positives, clean tail positives, and noisy positives at an early training epoch. Figure~\ref{fig:loss_ditribution} shows that clean tail positives can overlap with, and in this snapshot receive larger losses than, noisy samples. This suggests a plausible failure mode of small-loss denoising: some clean-but-hard tail signals may be down-weighted together with noisy interactions. Figure~\ref{fig:loss_ditribution} should be interpreted only as a diagnostic example, not as a universal verification across all datasets, backbones, or training epochs. Our subsequent theorem is conditional: it characterizes what happens when such a popularity-dependent loss pattern occurs.

\noindent\textbf{Diagnostic takeaway.} Large loss values need not correspond only to noisy interactions; in sparse tail regions, they may also arise from clean but hard positives.

To formalize this mechanism, we define item popularity as
\begin{equation}
\operatorname{pop}(i)
\;=\;
\mathbb{E}\!\left[\sum_{u} \mathbb{I}\!\left(Y_{ui}=1\right)\right]
=
\sum_{u} \mathbb{E}\!\left[\mathbb{I}\!\left(Y_{ui}=1\right)\right].
\label{eq:definition_pop}
\end{equation}
Let $\tau$ be a popularity threshold. We define the tail set as
$\mathcal{T}=\{(u,i)\mid \operatorname{pop}(i)\le \tau\}$ and the head set as $\mathcal{H}=\{(u,i)\mid \operatorname{pop}(i)>\tau\}$.
We consider the following condition.

\noindent\textbf{Condition 1 (Popularity-dependent loss pattern).}
Tail positives have larger positive losses than head positives in the sense that
\begin{equation}
\mathbb{E}_{(u,i)\in\mathcal{T}}\!\left[l^{+}(u,i)\right]
>
\mathbb{E}_{(u,i)\in\mathcal{H}}\!\left[l^{+}(u,i)\right].
\label{eq:assumption_2}
\end{equation}
For the cleanest theoretical statement, we use a distributional version of this condition in the proof: the positive-loss distribution on the tail is right-shifted relative to that on the head. This is a sufficient condition, not a necessary condition, and it should not be read as a universal empirical claim. Instead, it specifies the regime under which the following analysis applies.

\subsection{Effective Head--Tail Signal Ratio}
We next quantify how a denoising weight reallocates positive supervision between head and tail items.

\begin{definition}[Effective head--tail signal ratio]
Let $D^{+}=\{(u,i)\mid \bar{Y}_{ui}=1\}$ and sample $(U,I)$ uniformly from $D^{+}$. Define $H=\{(u,i)\in D^{+}\mid \operatorname{pop}(i)>\tau\}$ and $T=\{(u,i)\in D^{+}\mid \operatorname{pop}(i)\le\tau\}$. For any weighting scheme $w(\cdot,\cdot)$ applied to observed positives, define the effective head--tail signal ratio as
\begin{equation}
\mathcal{B}
:=
\frac{
\mathbb{E}\!\left[w(U,I)\mid (U,I)\in H\right]\cdot \mathbb{P}\!\left((U,I)\in H\right)
}{
\mathbb{E}\!\left[w(U,I)\mid (U,I)\in T\right]\cdot \mathbb{P}\!\left((U,I)\in T\right)
}.
\label{eq:definition_bias_intensity}
\end{equation}
\end{definition}

Here, $\mathbb{E}[w(U,I)\mid (U,I)\in H]$ and $\mathbb{E}[w(U,I)\mid (U,I)\in T]$ are the average weights assigned to head and tail observed positives, while $\mathbb{P}((U,I)\in H)$ and $\mathbb{P}((U,I)\in T)$ capture their prevalence in the observed positive set. Thus, $\mathcal{B}$ measures how much effective positive training signal is allocated to head items relative to tail items.

\begin{proposition}[Conditional Head--Tail Signal Reallocation]
Assume that the denoising weight is determined by a non-increasing function of the positive loss,
\begin{equation}
w(u,i)=\psi\!\left(l^{+}(u,i)\right),
\qquad \psi'(\cdot)\le 0,
\label{eq:definition_w}
\end{equation}
and assume that the positive-loss distribution on $T$ is right-shifted relative to that on $H$. Let $\mathcal{B}_{\mathrm{denoise}}$ and $\mathcal{B}_{\mathrm{ERM}}$ denote the effective head--tail signal ratios of the denoising method and ERM, respectively. Then
\begin{equation}
n_{\mathrm{denoise}}
:=
\frac{\mathcal{B}_{\mathrm{denoise}}}{\mathcal{B}_{\mathrm{ERM}}} > 1.
\label{eq:proposition_1}
\end{equation}
\end{proposition}

The proof is provided in Appendix~A. Proposition~1 should be read as a sufficient-condition result. It does not state that every denoising method always worsens popularity bias. Rather, it characterizes a mechanism that arises when two conditions hold: denoising weights decrease with loss, and tail positives tend to have larger losses than head positives. In this regime, denoising reduces the average weight of tail positives more strongly than that of head positives, thereby increasing the effective head--tail signal ratio relative to ERM. This provides a formal explanation of how loss-based denoising can interact with popularity bias.

\section{Popularity-Aware Denoising}

The conditional analysis above suggests a potential failure mode of uniform loss-based denoising: when tail positives tend to incur larger losses, applying the same denoising strength across all items may over-suppress clean tail signals. We therefore propose \emph{Popularity-Aware Denoising} (PAD), a simple plug-in correction that modulates the strength of a base denoiser by item popularity.

\subsection{Popularity-Aware Denoising (PAD)}

Recall that the goal of denoising is to approximate
\begin{equation}
w(u,i)
=
\mathbb{P}\!\left(
Y^{*}_{u,i}=1 \mid \bar{Y}_{u,i}=1,\, u,\, i
\right).
\label{eq:w_eq_p}
\end{equation}
Given any base denoising weight $w(u,i)$, PAD constructs a popularity-gated weight:
\begin{equation}
w_{\mathrm{PAD}}(u, i)=\left(1-s_i\right)+s_i\, w(u, i),
\label{eq:w_pad}
\end{equation}
where $s_i\in[0,1]$ is the popularity gating coefficient of item $i$. This design has two behaviors:
\begin{itemize}[leftmargin=*]
\item \textbf{Head items:} When $s_i\approx 1$, we have $w_{\mathrm{PAD}}(u,i)\approx w(u,i)$. PAD therefore preserves the ability of the base denoiser to suppress unreliable interactions in highly exposed regions.
\item \textbf{Tail items:} When $s_i\approx 0$, we have $w_{\mathrm{PAD}}(u,i)\approx 1$. PAD therefore weakens denoising on tail positives and avoids treating sparse clean-but-hard signals as noise solely because they are hard to fit.
\end{itemize}

Eq.~\eqref{eq:w_pad} defines an implementation-agnostic plug-in framework. It does not prescribe a particular denoising estimator; instead, it controls how strongly the base estimator is trusted as a function of item popularity. In this sense, PAD is a correction to uniform denoising rather than a replacement for existing denoising methods.

In our implementation, we instantiate the base denoising weight with a lightweight RCE-style reweighting \cite{wang2021denoising}:
\begin{equation}
w(u,i)=\hat{y}_{ui}^{\alpha}, \qquad \alpha\in[0,1],
\end{equation}
where $\hat{y}_{ui}\in(0,1)$ is the model prediction and $\alpha$ controls denoising strength. For a positive interaction, the BCE loss is $l^{+}(u,i)=-\log \hat{y}_{ui}$, so $\hat{y}_{ui}^{\alpha}=\exp(-\alpha l^{+}(u,i))$ is a monotone mapping that assigns smaller weights to larger-loss positives.

For the popularity gate, we use normalized item popularity:
\begin{equation}
\tilde{w}(u,i;\alpha,\eta)=\bigl(1-s_i(\eta)\bigr)+s_i(\eta)\,\hat{y}_{ui}^{\alpha}, \quad
s_i(\eta)=\left(\frac{\operatorname{pop}(i)}{\max_{j\in\mathcal{I}}\operatorname{pop}(j)}\right)^{\eta},
\label{eq:pad_instantiation}
\end{equation}
where $\eta\ge 0$ controls the strength of popularity-aware compensation. When $\eta\to 0$, most $s_i\to 1$ and PAD reduces to the base denoising paradigm. As $\eta$ increases, non-popular items receive smaller $s_i$ values, causing PAD to preserve more gradient contribution from tail positives.

The resulting training objective is
\begin{equation}
\mathcal L_{\mathrm{PAD}}(\bar{\mathcal D})
=
\frac{1}{|\bar{\mathcal D}|}\sum_{(u,i)\in \bar{\mathcal D}}
\tilde{w}(u,i;\alpha,\eta)\cdot \mathcal L_{\mathrm{BCE}}(u,i).
\label{eq:pad_objective}
\end{equation}
Intuitively, PAD keeps denoising active on head items while shrinking the denoising weight toward ERM on tail items. In practice, we also apply the weighting to negative samples in a symmetric manner so that the overall loss is not dominated by negatives. Moreover, to mitigate the distribution mismatch between validation and test sets \cite{zeng2025we}, we evaluate the validation criterion on a subset of low-loss validation samples for model selection.

\subsection{Discussion}

\subsubsection{PAD Reduces Conditional Signal Imbalance}
Under the same sufficient conditions as Proposition~1, PAD reduces the additional head--tail signal imbalance induced by monotone loss-based denoising.

\begin{proposition}[PAD Reduces Conditional Signal Imbalance]
Let $n_{\mathrm{PAD}} := \mathcal{B}_{\mathrm{PAD}} / \mathcal{B}_{\mathrm{ERM}}$ denote the head--tail signal ratio of PAD relative to ERM. Under the conditions of Proposition~1 and the PAD weighting rule in Eq.~\eqref{eq:w_pad},
\begin{equation}
n_{\mathrm{PAD}} < n_{\mathrm{denoise}}.
\label{eq:proposition_2}
\end{equation}
\end{proposition}

The proof is provided in Appendix~B. Proposition~2 shows that PAD reduces the additional head--tail signal imbalance caused by the base denoiser in the conditional regime analyzed above. The intuition is that PAD moves tail weights closer to $1$ more strongly than it moves head weights, thereby compensating for the over-suppression of tail positives.

\subsubsection{Collaborative-Signal-Only Hard-Positive Preservation}
PAD can also be interpreted as a collaborative-signal-only mechanism for preserving hard positives. Consider the classical two-stage generative process under the MNAR assumption \cite{schnabel2016recommendations, saito2020doubly, wang2019doubly}. For any $(u,i)$, exposure first occurs as
\begin{equation}
O_{ui} \sim \operatorname{Bernoulli}\!\left(\theta_{ui}\right),
\label{eq:bernoulli}
\end{equation}
followed by the generation of an interaction according to the user's true preference. The observed implicit positive feedback may contain false-positive noise induced by presentation effects, mis-clicks, or other exposure-related factors.

The popularity gate can be viewed as a coarse approximation of exposure propensity:
\begin{equation}
s_i \approx \mathbb{E}_{u}\!\left[\theta_{ui}\right].
\label{eq:propensity}
\end{equation}
When exposure propensity is high, observed positives are more likely to be affected by exposure-induced artifacts, so PAD relies more on the denoiser. When exposure propensity is low, an observed positive may reflect stronger intent, and aggressive small-loss denoising is more likely to confuse hard positives with noise. PAD therefore shrinks the denoising weight back toward $1$ in this region.

Unlike hard-positive mining methods that rely on textual or multimodal evidence \cite{wang2025unleashing, song2025hard}, PAD uses only collaborative signals and item popularity. It is therefore broadly applicable in settings where content features are unavailable.

\section{Experiments}

In this section, we evaluate the denoising--popularity interaction and the effect of PAD using the existing experimental setting. We focus on the following research questions:
\begin{itemize}[leftmargin=*]
\item \textbf{RQ1}: How does loss-based denoising affect popularity-related outcome diversity?
\item \textbf{RQ2}: Can PAD improve the accuracy--diversity tradeoff over representative denoising baselines?
\item \textbf{RQ3}: How does PAD behave under different denoising strengths and learning objectives?
\end{itemize}

\subsection{Experimental Setup}

\subsubsection{Datasets}

To ensure a fair evaluation, we conduct experiments on the three datasets that are most frequently adopted in the denoising recommendation: MovieLens, Amazon-book and Yelp. 

\begin{itemize}[leftmargin=*]
\item \textbf{MovieLens} \cite{harper2015movielens}: This dataset consists of movie rating records collected from the MovieLens platform. We use the MovieLens-100k version. Following \cite{wang2021denoising}, interactions with ratings lower than 5 are treated as false-positive feedback.
\item \textbf{Amazon-book}:  This dataset is derived from the Amazon Reviews collection \cite{he2016ups} and records users’ book purchase behaviors along with their rating scores. Interactions with ratings lower than 3 are treated as false-positive feedback.
\item \textbf{Yelp}
: This is a user review dataset from the catering industry. Consistent with the setting in \cite{wang2021denoising}, ratings below 3 are considered false-positive interactions.
\end{itemize}

In line with existing denoising recommendation methods, false-positive interactions are retained in both the training and validation sets, whereas only clean interactions are kept in the test set. The basic statistics of these datasets are summarized in Table~\ref{table-dataset}.

\begin{table}[t]
\centering
\caption{Statistics of the experimental datasets.}
\label{table-dataset}
\resizebox{1.0\columnwidth}{!}{%
\setlength{\aboverulesep}{0pt}
\setlength{\belowrulesep}{0.45pt} 
\begin{tabular}{lrrrr}
\toprule
\textbf{Dataset} & \textbf{\#Users} & \textbf{\#Items} & \textbf{\#Interactions} & \textbf{Sparsity} \\
\midrule
MovieLens & 943 & 1,682 & 100,000 & 93.695\% \\
Amazon-Book & 79,848 & 29,599 & 1,001,486 & 99.958\% \\
Yelp & 45,548 & 57,396 & 1,672,520 & 99.959\% \\
\bottomrule
\end{tabular}%
}
\end{table}

\begin{figure*}[t]
  \centering
  \includegraphics[width=0.99\textwidth]{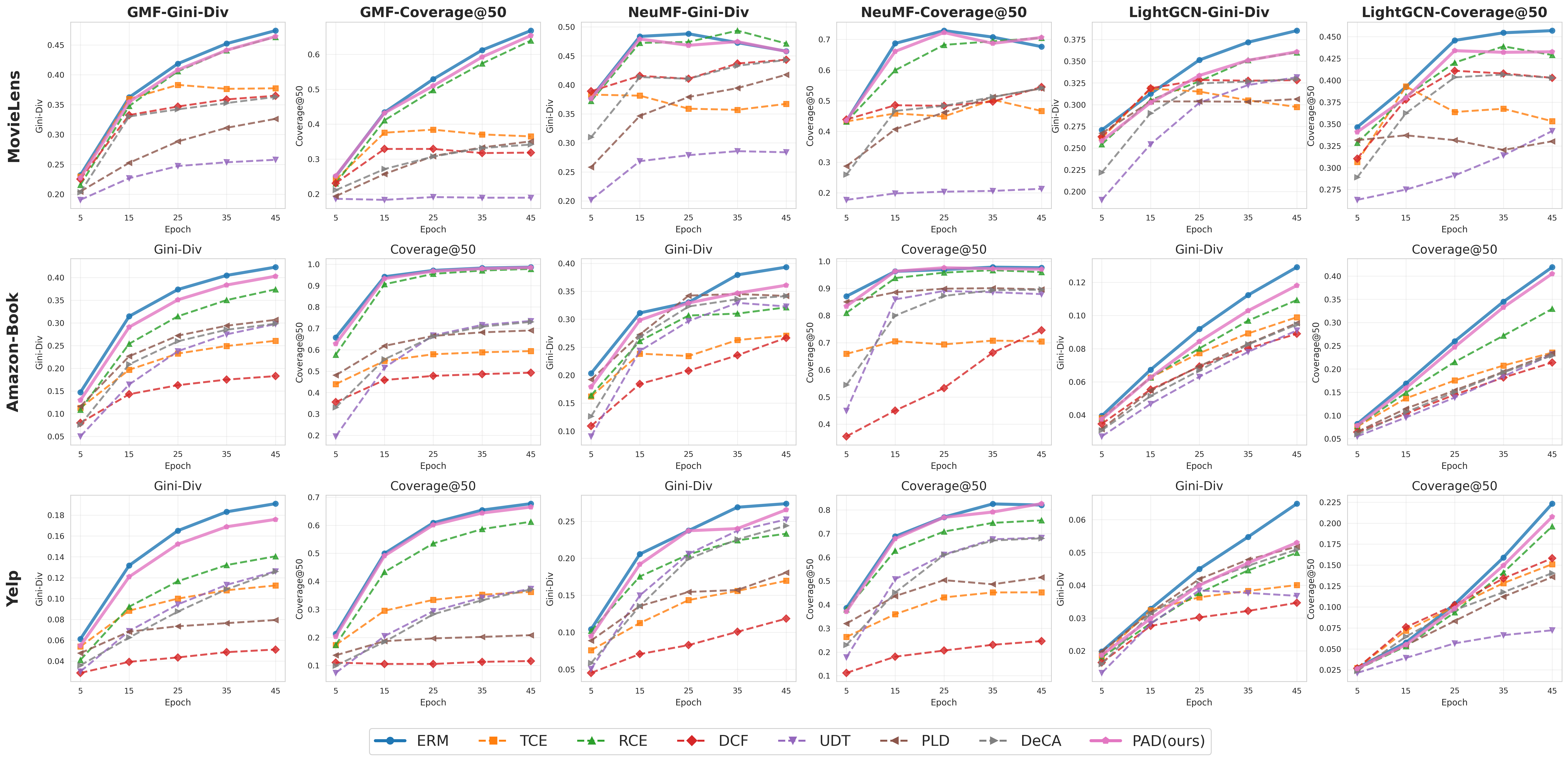}
  \caption{Epoch-wise evolution of diversity metrics (Gini-Div and Coverage@50; higher is better) under ERM (no denoising), representative denoising baselines, and PAD (ours) on MovieLens, Amazon-Book, and Yelp with GMF, NeuMF, and LightGCN.}
  \label{fig:diversity}
\end{figure*}

\subsubsection{Baselines}

In the denoising recommendation setting, we initially considered three representative models as baselines, including GMF, NeuMF and LightGCN:

\begin{itemize}[leftmargin=*]
\item \textbf{GMF} \cite{he2017neural}: an extension of matrix factorization that models user–item interactions through element-wise multiplication.
\item \textbf{NeuMF} \cite{wu2016collaborative}: integrates GMF with a multilayer perceptron to jointly capture non-linear user–item interaction patterns.
\item \textbf{LightGCN} \cite{he2017neural}: a graph-based model that learns user and item representations via linear message propagation.
\end{itemize}


We compare PAD with several representative baselines. We treat ERM, i.e., empirical risk minimization without denoising, as a no-denoising reference rather than as a denoising baseline. This distinction is important because our goal is not to claim that denoising is always preferable to ERM, but to examine how denoising can be made less popularity-biased when it is applied. R-CE \cite{wang2021denoising} re-weights user–item interactions based on their cross-entropy losses, assigning lower weights to high-loss samples. T-CE \cite{wang2021denoising} filters out interactions whose losses exceed a predefined threshold during training. DeCA \cite{wang2022learning} detects noisy interactions by measuring prediction disagreement between two models. DCF \cite{he2024double} alleviates sparsity by re-labeling highly deterministic noisy interactions instead of discarding them. \cite{zhang2025personalized} proposes a personalized approach to re-weighting user interactions. UDT \cite{chua2024unified} models user behavior as a two-stage process and adaptively adjusts interaction weights based on uncertainty patterns. These methods represent commonly used and recent denoising strategies and serve as our primary denoising points of comparison. For fairness \footnote{Two denoising methods are not included in our comparisons. We exclude \cite{gao2022self} due to its high computational cost; this limitation is also noted in \cite{chua2024unified}. We also exclude \cite{wang2023efficient} because its implementation may suffer from test-set data leakage, as noted by \cite{zeng2025we}.}, we do not include denoising approaches \cite{wang2025unleashing, song2025hard} that rely on large language models with textual information for data augmentation.


\subsubsection{Evaluation Metrics}
For recommendation accuracy, we evaluate the top-$N$ recommendations using two commonly adopted metrics: Recall@K and NDCG@K, where higher values indicate better performance. To assess popularity bias, we use two aggregate-level diversity metrics, Coverage@K and Gini-Div. Intuitively, Coverage@K reflects how many distinct items are exposed across all users’ Top-$K$ lists, while Gini-Div captures how evenly item exposure is distributed (i.e., less concentration on a few popular items). For both metrics, larger values indicate higher diversity and thus a fairer (less popularity-biased) recommendation outcome.

\begin{table*}[t]
\centering
\caption{Comparison with representative denoising methods using GMF, NeuMF, and LightGCN as backbones. The best performance for each backbone is in \textbf{bold}, and the second-best is \underline{underlined}. We use R to stand for Recall and N for NDCG.}
\label{tab:debiasing_comparison}
\renewcommand{\arraystretch}{0.85}
\resizebox{\textwidth}{!}{%
\begin{tabular}{l|cccc|cccc|cccc}
\toprule

& \multicolumn{4}{c|}{\textbf{MovieLens}} & \multicolumn{4}{c|}{\textbf{Amazon-Book}} & \multicolumn{4}{c}{\textbf{Yelp}} \\
\textbf{Method} & \textbf{R@50} & \textbf{N@50} & \textbf{R@100} & \textbf{N@100} & \textbf{R@50} & \textbf{N@50} & \textbf{R@100} & \textbf{N@100} & \textbf{R@50} & \textbf{N@50} & \textbf{R@100} & \textbf{N@100} \\
\midrule
GMF-ERM & 0.1998 & 0.0947 & 0.2703 & 0.1130 & 0.1222 & 0.0453 & 0.1822 & 0.0567 & 0.0857 & 0.0359 & 0.1361 & 0.0473 \\
GMF-RCE & 0.1985 & 0.0974 & 0.2764 & 0.1178 & 0.1231 & 0.0452 & 0.1824 & 0.0566 & 0.0868 & \underline{0.0367} & 0.1379 & 0.0482 \\
GMF-TCE & \underline{0.2065} & \underline{0.1020} & \underline{0.2769} & \underline{0.1203} & \underline{0.1294} & \underline{0.0480} & \underline{0.1936} & \underline{0.0602} & \underline{0.0885} & 0.0356 & \underline{0.1460} & \underline{0.0486} \\
GMF-DeCA & 0.1616 & 0.0862 & 0.2382 & 0.1059 & 0.0874 & 0.0336 & 0.1284 & 0.0416 & 0.0677 & 0.0292 & 0.1063 & 0.0380 \\
GMF-DCF & 0.1579 & 0.0842 & 0.2344 & 0.1035 & 0.0895 & 0.0343 & 0.1315 & 0.0425 & 0.0667 & 0.0287 & 0.1047 & 0.0375 \\
GMF-PLD & 0.1862 & 0.0925 & 0.2583 & 0.1112 & 0.1222 & 0.0453 & 0.1822 & 0.0567 & 0.0857 & 0.0359 & 0.1361 & 0.0473 \\
GMF-UDT & 0.1630 & 0.0898 & 0.2327 & 0.1079 & 0.0981 & 0.0375 & 0.1445 & 0.0464 & 0.0816 & 0.0348 & 0.1283 & 0.0453 \\
\textbf{GMF-PAD} & \textbf{0.2085} & \textbf{0.1101} & \textbf{0.2834} & \textbf{0.1293} & \textbf{0.1369} & \textbf{0.0499} & \textbf{0.2040} & \textbf{0.0628} & \textbf{0.0971} & \textbf{0.0389} & \textbf{0.1570} & \textbf{0.0523} \\
\rowcolor{lightgray}
Improvement & +1.0\% & +8.0\% & +2.4\% & +7.5\% & +5.8\% & +4.1\% & +5.4\% & +4.3\% & +9.7\% & +6.0\% & +7.5\% & +7.8\% \\
\midrule
\midrule
NeuMF-ERM & 0.1974 & 0.0985 & 0.2802 & 0.1193 & 0.1104 & 0.0399 & 0.1691 & 0.0511 & 0.0728 & 0.0293 & 0.1216 & 0.0402 \\
NeuMF-RCE & 0.1989 & \underline{0.0999} & \underline{0.2863} & \underline{0.1212} & 0.1086 & 0.0391 & 0.1671 & 0.0503 & \underline{0.0772} & 0.0315 & 0.1275 & \underline{0.0429} \\
NeuMF-TCE & 0.1965 & 0.0971 & 0.2722 & 0.1164 & \underline{0.1192} & \underline{0.0431} & \underline{0.1805} & \underline{0.0548} & 0.0748 & 0.0292 & \underline{0.1292} & 0.0414 \\
NeuMF-DeCA & 0.1510 & 0.0755 & 0.2235 & 0.0935 & 0.0460 & 0.0157 & 0.0719 & 0.0206 & 0.0707 & 0.0288 & 0.1157 & 0.0394 \\
NeuMF-DCF & 0.1551 & 0.0775 & 0.2296 & 0.0961 & 0.0448 & 0.0153 & 0.0701 & 0.0201 & 0.0716 & 0.0296 & 0.1171 & 0.0398 \\
NeuMF-PLD & \underline{0.2027} & 0.0999 & 0.2793 & 0.1194 & 0.1110 & 0.0399 & 0.1695 & 0.0511 & 0.0728 & 0.0293 & 0.1216 & 0.0402 \\
NeuMF-UDT & 0.1575 & 0.0786 & 0.2341 & 0.0973 & 0.0735 & 0.0263 & 0.1140 & 0.0342 & 0.0760 & \underline{0.0319} & 0.1227 & 0.0424 \\
\textbf{NeuMF-PAD} & \textbf{0.2081} & \textbf{0.1100} & \textbf{0.2865} & \textbf{0.1298} & \textbf{0.1279} & \textbf{0.0462} & \textbf{0.1927} & \textbf{0.0586} & \textbf{0.0802} & \textbf{0.0320} & \textbf{0.1349} & \textbf{0.0442} \\
\rowcolor{lightgray}
Improvement & +2.7\% & +10.1\% & +0.1\% & +7.1\% & +7.3\% & +7.2\% & +6.7\% & +6.8\% & +3.8\% & +0.2\% & +4.4\% & +3.2\% \\
\midrule
\midrule
LightGCN-ERM & 0.1938 & 0.1000 & \underline{0.2746} & \underline{0.1204} & \textbf{0.1766} & \textbf{0.0660} & \underline{0.2579} & \textbf{0.0814} & \textbf{0.1215} & \textbf{0.0520} & \textbf{0.1888} & \textbf{0.0671} \\
LightGCN-RCE & \underline{0.1995} & 0.1001 & 0.2742 & 0.1197 & 0.1394 & 0.0516 & 0.2067 & 0.0643 & 0.0932 & 0.0393 & 0.1450 & 0.0510 \\
LightGCN-TCE & 0.1961 & \underline{0.1002} & 0.2708 & 0.1195 & 0.1351 & 0.0499 & 0.2018 & 0.0625 & 0.0989 & 0.0425 & 0.1535 & 0.0549 \\
LightGCN-DeCA & 0.1534 & 0.0817 & 0.2275 & 0.1002 & 0.1372 & 0.0507 & 0.2049 & 0.0635 & 0.0719 & 0.0308 & 0.1126 & 0.0400 \\
LightGCN-DCF & 0.1882 & 0.0926 & 0.2527 & 0.1096 & 0.1411 & 0.0532 & 0.2069 & 0.0658 & 0.0737 & 0.0315 & 0.1154 & 0.0409 \\
LightGCN-PLD & 0.1944 & 0.0975 & 0.2656 & 0.1157 & 0.1731 & \underline{0.0650} & 0.2497 & 0.0796 & 0.1059 & 0.0452 & 0.1654 & 0.0586 \\
LightGCN-UDT & 0.1518 & 0.0808 & 0.2250 & 0.0991 & 0.1459 & 0.0539 & 0.2133 & 0.0667 & 0.0945 & 0.0400 & 0.1470 & 0.0518 \\
\textbf{LightGCN-PAD} & \textbf{0.2018} & \textbf{0.1028} & \textbf{0.2795} & \textbf{0.1228} & \underline{0.1749} & 0.0644 & \textbf{0.2579} & \underline{0.0801} & \underline{0.1140} & \underline{0.0475} & \underline{0.1797} & \underline{0.0622} \\
\rowcolor{lightgray}
Improvement & +1.2\% & +2.7\% & +1.8\% & +2.0\% & -1.0\% & -2.4\% & +0.0\% & -1.6\% & -6.2\% & -8.7\% & -4.8\% & -7.4\% \\
\bottomrule
\end{tabular}%
}
\label{tabletable}
\end{table*}

\subsubsection{Implementation Details}

For all models, we fix the embedding dimensionality to 32, use a batch size of 1024, and set the learning rate to 1e-3, with training carried out using the Adam optimizer. For baseline methods, the remaining hyperparameters are adopted from the optimal configurations reported in their original publications. Since denoising methods can exhibit sensitivity to dataset characteristics, we perform limited hyperparameter tuning in cases where the authors do not explicitly specify parameter settings, ensuring a fair evaluation across different datasets. In the proposed PAD method, there are two hyperparameters, $\alpha$ and $\eta$. We search $\alpha$ in the range $[0.1, 0.5]$ and $\eta$ in the range $[0.3, 0.7]$. Based on our experiments, we primarily set $\alpha = 0.2$ and $\eta = 0.5$. For model selection, we use the lowest-loss $80\%$ of validation examples.

\subsection{Outcome Diversity under Denoising (RQ1)}
To examine how loss-based denoising affects popularity-related outcomes, we evaluate two diversity-oriented metrics: Gini-Div and Coverage@50, where higher values indicate more even and broader item exposure. Figure~\ref{fig:diversity} reports epoch-wise trajectories under ERM, representative denoising baselines, and PAD. We plot trajectories rather than a single checkpoint because different methods have different convergence dynamics; the full curves better reveal whether a method preserves or weakens diversity throughout training.

Figure~\ref{fig:diversity} shows that representative denoising baselines often yield lower Coverage@50 and Gini-Div than ERM, especially as training proceeds. This pattern is consistent with the conditional mechanism analyzed in Section~3: when high-loss positives are down-weighted, supervision from hard tail regions may be weakened, leading to more concentrated recommendation outcomes. PAD generally mitigates this drop and often yields higher diversity than the corresponding uniform denoising baselines. The magnitude of the effect varies across datasets and backbones, which supports our conditional view rather than a universal claim.

The role of Figure~\ref{fig:diversity} is complementary to that of Figure~\ref{fig:loss_ditribution}. Figure~\ref{fig:loss_ditribution} illustrates a microscopic loss-level failure mode in one representative setting, whereas Figure~\ref{fig:diversity} examines downstream outcome-level diversity across multiple datasets and backbones. Together, they suggest that popularity-aware modulation can reduce the diversity loss introduced by aggressive denoising.

\subsection{Accuracy--Diversity Tradeoff with PAD (RQ2)}
Table~\ref{tabletable} reports ranking performance on three datasets and three backbones. We interpret the results along two axes: comparison with representative denoising baselines and comparison with the no-denoising ERM reference.

On MF-style backbones, PAD is particularly effective. For GMF and NeuMF, PAD achieves the best performance in most metrics and datasets, indicating that popularity-aware denoising is useful when sparse tail signals are vulnerable to over-suppression. For example, on MovieLens, PAD improves GMF's NDCG@50 over the strongest competing method, and on Amazon-Book it improves NeuMF's Recall@50 over the strongest competing baseline. These gains suggest that PAD can preserve informative hard positives while retaining the robustness benefit of denoising.

On LightGCN, ERM is a strong reference and sometimes outperforms PAD, especially on Amazon-Book and Yelp. We view this as an important boundary case rather than a contradiction. Graph propagation can exploit multi-hop neighborhood structure and may partially smooth sparse or noisy signals, thereby reducing the need for explicit denoising; this is also consistent with the discussion in \cite{zeng2025we}. Nevertheless, PAD remains competitive and is often preferable to applying uniform loss-based denoising. Overall, the table supports a more nuanced conclusion: PAD improves the behavior of denoising methods, especially on MF-style backbones, but denoising is not universally superior to ERM.

\subsection{Compatibility \& Sensitivity Analysis (RQ3)}

\begin{figure}[]
  \centering
  \includegraphics[width=0.48\textwidth]{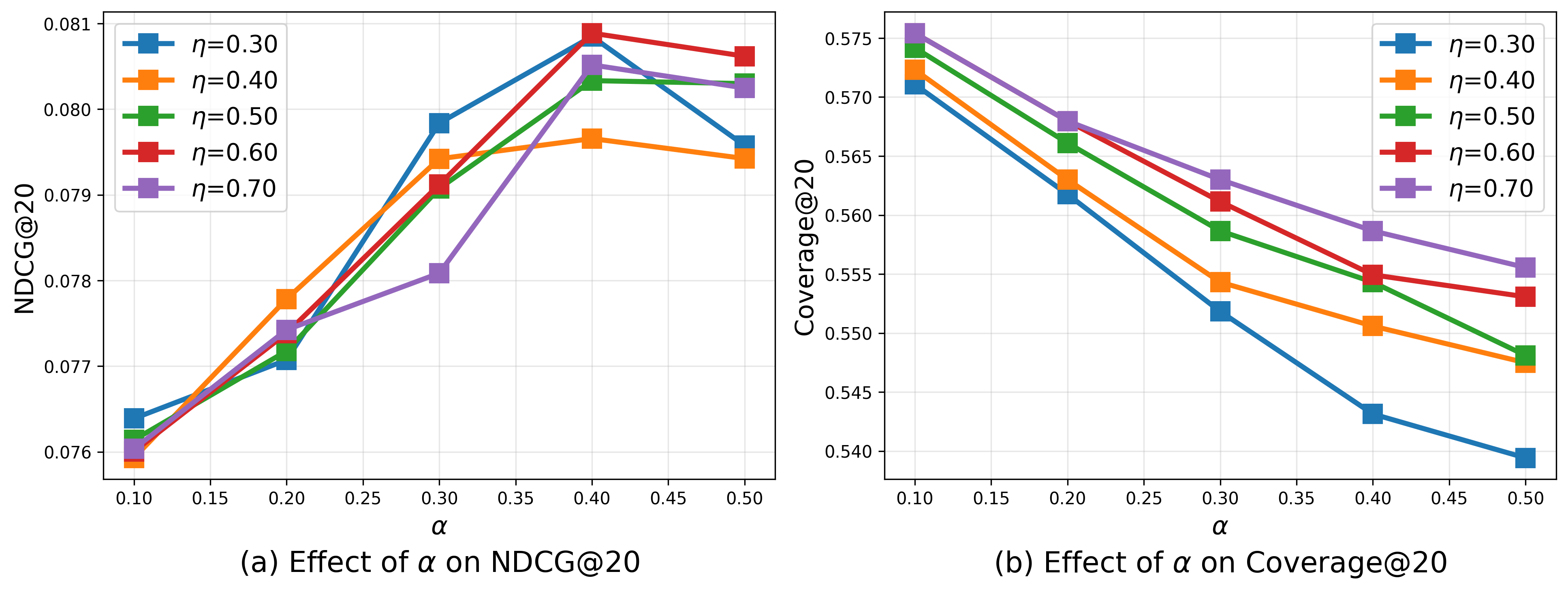}
  \caption{Sensitivity analysis of hyperparameters $\alpha$ and $\eta$ on MovieLens with the GMF backbone.}
  \label{fig:hyper}
\end{figure}

\noindent\textbf{Hyperparameter sensitivity.}
We investigate the sensitivity of PAD to its two hyperparameters, $\alpha$ and $\eta$. Here, $\alpha$ controls the strength of the base denoising effect, while $\eta$ controls the popularity-aware compensation. Figure~\ref{fig:hyper} reports their influence on MovieLens with the GMF backbone.

The sensitivity curves provide direct evidence for the mechanism behind PAD. As $\alpha$ increases from small to moderate values, NDCG@20 improves under several $\eta$ settings, suggesting that moderate denoising can suppress noisy supervision and improve ranking. However, stronger denoising also tends to reduce Coverage@20, reflecting the accuracy--coverage tradeoff induced by small-loss weighting. By contrast, increasing $\eta$ generally recovers coverage under the same $\alpha$ while affecting NDCG more mildly. This is consistent with the role of popularity-aware compensation: $\eta$ counteracts excessive tail suppression without fully abandoning denoising on head items.

\noindent\textbf{Compatibility with BPR.}
Since our main derivation uses a point-wise BCE objective, we also conduct a limited compatibility check with BPR on Yelp. BPR is a representative pair-wise objective and differs from BCE, so it provides a useful test of whether the PAD gating idea is tied to a particular point-wise loss.

\begin{table}[t]
    \renewcommand\arraystretch{0.9}
\setlength{\abovecaptionskip}{0.1cm}
\setlength{\belowcaptionskip}{-0.2cm}
    \caption{Recall@50 comparison with BPR loss on Yelp dataset.}
    \tabcolsep=2pt
    \begin{tabular}{lcccccc}
    \toprule
    \multirow{2}{*}{\textbf{Method}}&\multicolumn{2}{c}{\textbf{GMF}} &\multicolumn{2}{c}{\textbf{NeuMF}} &\multicolumn{2}{c}{\textbf{LightGCN}}\\ \cmidrule(r){2-3}\cmidrule(r){4-5}\cmidrule(r){6-7}&{+BCE}&+BPR&+BCE&+BPR&+BCE&+BPR \\
    \midrule 
    ERM&0.0857&0.1040&0.0728&0.0896&\textbf{0.1215}&\underline{0.1381}\\
    RCE&0.0868&0.1017&\underline{0.0772}&0.0995&0.0932&0.1314\\
    TCE&\underline{0.0885}&0.0847&0.0748&0.0464&0.0989&0.0770\\
    DeCA&0.0677&0.0840&0.0707&0.0460&0.0719&0.0703\\
    DCF&0.0667&\underline{0.1112}&0.0716&\underline{0.1010}&0.0737&0.0737\\
    PLD&0.0857&0.0825&0.0728&0.0773&0.1059&0.0893\\
    UDT&0.0816&0.0848&0.0760&0.0763&0.0945&0.0942\\
    PAD&\textbf{0.0971}&\textbf{0.1121}&\textbf{0.0802}&\textbf{0.1061}&\underline{0.1140}&\textbf{0.1382}\\
    \rowcolor{lightgray} 
    Improvement&+9.7\%&+0.8\%&+3.8\%&+5.0\%&-6.6\%&+0.1\%\\
    \bottomrule    
    \end{tabular}\label{performance_loss_yelp_recall_50}
    \vspace{-5pt}
    \label{table_BPR}
    \end{table}

Table~\ref{table_BPR} shows that PAD remains effective on GMF and NeuMF under BPR and is comparable to ERM on LightGCN. The margin over LightGCN+ERM is very small, so this result should be interpreted as a compatibility check rather than broad proof of objective-level generality. Overall, the BPR results suggest that popularity-aware gating can be instantiated beyond BCE, while a broader study of contrastive objectives and stronger modern backbones remains an important direction for future work.

\subsection{Scope and Limitations}
Our study focuses on understanding and mitigating a specific failure mode of loss-based denoising: popularity-dependent suppression of positive signals. The theoretical results are conditional and should be interpreted as characterizing a sufficient regime under which denoising increases the effective head--tail signal ratio. Empirically, our results are based on point estimates from representative datasets, backbones, denoising baselines, and a limited BPR compatibility check. We do not claim that denoising always outperforms ERM or that PAD universally improves all backbones. In fact, the strong performance of LightGCN+ERM suggests that graph-based propagation can sometimes reduce the need for explicit denoising. A broader investigation with additional modern self-supervised backbones, contrastive objectives, and multi-seed statistical testing is an important direction for future work.

\section{Related Work}

\subsection{Denoising Recommendation}
Implicit feedback has attracted increasing attention in recommender systems due to its ease of collection, but its data quality is susceptible to various types of noise~\cite{chen2020bias, abdollahpouri2019managing}. Noisy positives can mislead recommendation models and cause them to fit incorrect user preference patterns~\cite{wang2021denoising, wang2023tutorial, tan2022partial, lee2021bootstrapping}. To address this issue, denoising recommendation methods have been proposed~\cite{zhang2023robust, wang2023tutorial}.

The core of denoising recommendation is to estimate $P(Y^{\ast}\mid Y,u,i)$. Since $Y^{\ast}$ is unobservable, most collaborative-filtering denoising methods rely on heuristic estimators, many of which use the small-loss principle either directly or indirectly. As a pioneering work, \cite{wang2021denoising} learns robust recommenders by dynamically down-weighting high-loss samples or discarding them during training. Subsequent methods extend this idea in different ways: \cite{he2024double} flips labels of potentially noisy interactions, \cite{chua2024unified} models user behavior through willingness and action phases, \cite{zhang2025personalized} uses user-specific loss distributions, and \cite{wang2022learning} identifies noise by prediction disagreement across models. Beyond collaborative signals, some recent methods use textual or multimodal information for denoising and hard-sample mining \cite{wang2025unleashing, song2025hard}, while others are tailored to specific recommendation scenarios or model structures \cite{tian2022learning, zhang2024ssdrec, zhang2022hierarchical,wu2025empowering}.

Different from these works, our focus is not to design a more complex noise detector. Instead, we study how loss-based denoising itself interacts with popularity-induced head--tail imbalance, and we propose a lightweight gate that can be applied on top of a base denoising weight.

\subsection{Popularity Bias and Exposure-Aware Learning}
Popularity bias is a long-standing issue in recommender systems: popular items tend to receive more exposure and feedback, which further increases their dominance in both learned representations and recommendation outcomes \cite{abdollahpouri2019popularity, abdollahpouri2019managing, abdollahpouri2020connection, ren2022mitigating, lin2025recommendation}. Prior work studies this issue from multiple perspectives, including biased exposure, evaluation mismatch, representation concentration, and aggregate-level diversity.

PAD is related in spirit to exposure-aware learning, but its goal is different from classical inverse-propensity-weighting style debiasing. IPW-style methods aim to correct biased observation probabilities by inversely weighting samples, which requires estimating exposure propensities. PAD instead modulates the strength of denoising weights to avoid over-suppressing tail positives. It should therefore be viewed as a correction to denoising, not as a general replacement for exposure-debiased learning.

\section{Conclusion}

This work revisits recommendation denoising through the lens of popularity bias. We show that loss-based denoising can interact with item popularity: when tail positives incur larger losses than head positives, monotone small-loss reweighting may disproportionately reduce effective tail supervision and increase the head--tail signal ratio. We formalize this conditional mechanism and propose PAD, a lightweight popularity-aware gating framework that mitigates excessive tail suppression by adapting denoising strength across head and tail items. Experiments show that PAD often improves over representative denoising baselines and provides favorable accuracy--diversity tradeoffs, especially on MF-style backbones, while also revealing boundary cases where ERM remains competitive. These findings suggest that denoising should be applied in a popularity-aware rather than uniform manner, and that future denoising recommenders should evaluate not only robustness and accuracy, but also how denoising reallocates supervision across popularity groups.



\clearpage

\bibliographystyle{ACM-Reference-Format}
\bibliography{sample-base}

@inproceedings{wang2021denoising,
  title={Denoising implicit feedback for recommendation},
  author={Wang, Wenjie and Feng, Fuli and He, Xiangnan and Nie, Liqiang and Chua, Tat-Seng},
  booktitle={Proceedings of the 14th ACM international conference on web search and data mining},
  pages={373--381},
  year={2021}
}

@inproceedings{wang2022learning,
  title={Learning robust recommenders through cross-model agreement},
  author={Wang, Yu and Xin, Xin and Meng, Zaiqiao and Jose, Joemon M and Feng, Fuli and He, Xiangnan},
  booktitle={Proceedings of the ACM web conference 2022},
  pages={2015--2025},
  year={2022}
}

@inproceedings{zhang2024ssdrec,
  title={Ssdrec: Self-augmented sequence denoising for sequential recommendation},
  author={Zhang, Chi and Han, Qilong and Chen, Rui and Zhao, Xiangyu and Tang, Peng and Song, Hongtao},
  booktitle={2024 IEEE 40th International Conference on Data Engineering (ICDE)},
  pages={803--815},
  year={2024},
  organization={IEEE}
}

@article{song2025hard,
  title={Hard vs. Noise: Resolving Hard-Noisy Sample Confusion in Recommender Systems via Large Language Models},
  author={Song, Tianrui and Chao, Wen-Shuo and Liu, Hao},
  journal={arXiv preprint arXiv:2511.07295},
  year={2025}
}

@inproceedings{wang2025unleashing,
  title={Unleashing the Power of Large Language Model for Denoising Recommendation},
  author={Wang, Shuyao and Zheng, Zhi and Sui, Yongduo and Xiong, Hui},
  booktitle={Proceedings of the ACM on Web Conference 2025},
  pages={252--263},
  year={2025}
}

@inproceedings{tian2022learning,
  title={Learning to denoise unreliable interactions for graph collaborative filtering},
  author={Tian, Changxin and Xie, Yuexiang and Li, Yaliang and Yang, Nan and Zhao, Wayne Xin},
  booktitle={Proceedings of the 45th international ACM SIGIR conference on research and development in information retrieval},
  pages={122--132},
  year={2022}
}

@inproceedings{zhang2022hierarchical,
  title={Hierarchical item inconsistency signal learning for sequence denoising in sequential recommendation},
  author={Zhang, Chi and Du, Yantong and Zhao, Xiangyu and Han, Qilong and Chen, Rui and Li, Li},
  booktitle={Proceedings of the 31st ACM international conference on information \& knowledge management},
  pages={2508--2518},
  year={2022}
}

@inproceedings{wu2025empowering,
  title={Empowering Denoising Sequential Recommendation with Large Language Model Embeddings},
  author={Wu, Tongzhou and Wang, Yuhao and Wang, Maolin and Zhang, Chi and Zhao, Xiangyu},
  booktitle={Proceedings of the 34th ACM International Conference on Information and Knowledge Management},
  pages={3427--3437},
  year={2025}
}

@inproceedings{saito2020doubly,
  title={Doubly robust estimator for ranking metrics with post-click conversions},
  author={Saito, Yuta},
  booktitle={Proceedings of the 14th ACM Conference on Recommender Systems},
  pages={92--100},
  year={2020}
}

@inproceedings{wang2019doubly,
  title={Doubly robust joint learning for recommendation on data missing not at random},
  author={Wang, Xiaojie and Zhang, Rui and Sun, Yu and Qi, Jianzhong},
  booktitle={International Conference on Machine Learning},
  pages={6638--6647},
  year={2019},
  organization={PMLR}
}

@inproceedings{schnabel2016recommendations,
  title={Recommendations as treatments: Debiasing learning and evaluation},
  author={Schnabel, Tobias and Swaminathan, Adith and Singh, Ashudeep and Chandak, Navin and Joachims, Thorsten},
  booktitle={international conference on machine learning},
  pages={1670--1679},
  year={2016},
  organization={PMLR}
}

@inproceedings{zeng2025we,
  title={Are We Really Making Recommendations Robust? Revisiting Model Evaluation for Denoising Recommendation},
  author={Zeng, Guohang and Lu, Jie and Zhang, Guangquan},
  booktitle={Proceedings of the Nineteenth ACM Conference on Recommender Systems},
  pages={706--715},
  year={2025}
}

@book{schutze2008introduction,
  title={Introduction to information retrieval},
  author={Sch{\"u}tze, Hinrich and Manning, Christopher D and Raghavan, Prabhakar},
  volume={39},
  year={2008},
  publisher={Cambridge University Press Cambridge}
}

@article{cossock2008statistical,
  title={Statistical analysis of Bayes optimal subset ranking},
  author={Cossock, David and Zhang, Tong},
  journal={IEEE Transactions on Information Theory},
  volume={54},
  number={11},
  pages={5140--5154},
  year={2008},
  publisher={IEEE}
}

@inproceedings{di2025theoretical,
  title={A theoretical analysis of recommendation loss functions under negative sampling},
  author={Di Teodoro, Giulia and Siciliano, Federico and Tonellotto, Nicola and Silvestri, Fabrizio},
  booktitle={2025 International Joint Conference on Neural Networks (IJCNN)},
  pages={1--9},
  year={2025},
  organization={IEEE}
}

@inproceedings{bruch2019analysis,
  title={An analysis of the softmax cross entropy loss for learning-to-rank with binary relevance},
  author={Bruch, Sebastian and Wang, Xuanhui and Bendersky, Michael and Najork, Marc},
  booktitle={Proceedings of the 2019 ACM SIGIR international conference on theory of information retrieval},
  pages={75--78},
  year={2019}
}

@article{chen2023bias,
  title={Bias and debias in recommender system: A survey and future directions},
  author={Chen, Jiawei and Dong, Hande and Wang, Xiang and Feng, Fuli and Wang, Meng and He, Xiangnan},
  journal={ACM Transactions on Information Systems},
  volume={41},
  number={3},
  pages={1--39},
  year={2023},
  publisher={ACM New York, NY}
}

@inproceedings{baeza2020bias,
  title={Bias in search and recommender systems},
  author={Baeza-Yates, Ricardo},
  booktitle={Proceedings of the 14th ACM conference on recommender systems},
  pages={2--2},
  year={2020}
}

@inproceedings{hu2008collaborative,
  title={Collaborative filtering for implicit feedback datasets},
  author={Hu, Yifan and Koren, Yehuda and Volinsky, Chris},
  booktitle={2008 Eighth IEEE international conference on data mining},
  pages={263--272},
  year={2008},
  organization={Ieee}
}

@article{rendle2012bpr,
  title={BPR: Bayesian personalized ranking from implicit feedback},
  author={Rendle, Steffen and Freudenthaler, Christoph and Gantner, Zeno and Schmidt-Thieme, Lars},
  journal={arXiv preprint arXiv:1205.2618},
  year={2012}
}

@inproceedings{joachims2017accurately,
  title={Accurately interpreting clickthrough data as implicit feedback},
  author={Joachims, Thorsten and Granka, Laura and Pan, Bing and Hembrooke, Helene and Gay, Geri},
  booktitle={Acm Sigir Forum},
  volume={51},
  number={1},
  pages={4--11},
  year={2017},
  organization={Acm New York, NY, USA}
}

@inproceedings{abdollahpouri2019popularity,
  title={Popularity bias in ranking and recommendation},
  author={Abdollahpouri, Himan},
  booktitle={Proceedings of the 2019 AAAI/ACM Conference on AI, Ethics, and Society},
  pages={529--530},
  year={2019}
}

@article{abdollahpouri2019managing,
  title={Managing popularity bias in recommender systems with personalized re-ranking},
  author={Abdollahpouri, Himan and Burke, Robin and Mobasher, Bamshad},
  journal={arXiv preprint arXiv:1901.07555},
  year={2019}
}

@inproceedings{abdollahpouri2020connection,
  title={The connection between popularity bias, calibration, and fairness in recommendation},
  author={Abdollahpouri, Himan and Mansoury, Masoud and Burke, Robin and Mobasher, Bamshad},
  booktitle={Proceedings of the 14th ACM conference on recommender systems},
  pages={726--731},
  year={2020}
}

@article{han2018co,
  title={Co-teaching: Robust training of deep neural networks with extremely noisy labels},
  author={Han, Bo and Yao, Quanming and Yu, Xingrui and Niu, Gang and Xu, Miao and Hu, Weihua and Tsang, Ivor and Sugiyama, Masashi},
  journal={Advances in neural information processing systems},
  volume={31},
  year={2018}
}

@inproceedings{yu2019does,
  title={How does disagreement help generalization against label corruption?},
  author={Yu, Xingrui and Han, Bo and Yao, Jiangchao and Niu, Gang and Tsang, Ivor and Sugiyama, Masashi},
  booktitle={International conference on machine learning},
  pages={7164--7173},
  year={2019},
  organization={PMLR}
}

@inproceedings{lin2025recommendation,
  title={How do recommendation models amplify popularity bias? An analysis from the spectral perspective},
  author={Lin, Siyi and Gao, Chongming and Chen, Jiawei and Zhou, Sheng and Hu, Binbin and Feng, Yan and Chen, Chun and Wang, Can},
  booktitle={Proceedings of the Eighteenth ACM International Conference on Web Search and Data Mining},
  pages={659--668},
  year={2025}
}

@inproceedings{ren2022mitigating,
  title={Mitigating popularity bias in recommendation with unbalanced interactions: A gradient perspective},
  author={Ren, Weijieying and Wang, Lei and Liu, Kunpeng and Guo, Ruocheng and Peng, Lim Ee and Fu, Yanjie},
  booktitle={2022 IEEE International Conference on Data Mining (ICDM)},
  pages={438--447},
  year={2022},
  organization={IEEE}
}

@inproceedings{yang2018unbiased,
  title={Unbiased offline recommender evaluation for missing-not-at-random implicit feedback},
  author={Yang, Longqi and Cui, Yin and Xuan, Yuan and Wang, Chenyang and Belongie, Serge and Estrin, Deborah},
  booktitle={Proceedings of the 12th ACM conference on recommender systems},
  pages={279--287},
  year={2018}
}

@article{tolomei2019you,
  title={You must have clicked on this ad by mistake! Data-driven identification of accidental clicks on mobile ads with applications to advertiser cost discounting and click-through rate prediction},
  author={Tolomei, Gabriele and Lalmas, Mounia and Farahat, Ayman and Haines, Andrew},
  journal={International Journal of Data Science and Analytics},
  volume={7},
  pages={53--66},
  year={2019},
  publisher={Springer}
}

@inproceedings{he2017neural,
  title={Neural collaborative filtering},
  author={He, Xiangnan and Liao, Lizi and Zhang, Hanwang and Nie, Liqiang and Hu, Xia and Chua, Tat-Seng},
  booktitle={Proceedings of the 26th international conference on world wide web},
  pages={173--182},
  year={2017}
}

@inproceedings{wu2016collaborative,
  title={Collaborative denoising auto-encoders for top-n recommender systems},
  author={Wu, Yao and DuBois, Christopher and Zheng, Alice X and Ester, Martin},
  booktitle={Proceedings of the ninth ACM international conference on web search and data mining},
  pages={153--162},
  year={2016}
}

@inproceedings{he2024double,
  title={Double correction framework for denoising recommendation},
  author={He, Zhuangzhuang and Wang, Yifan and Yang, Yonghui and Sun, Peijie and Wu, Le and Bai, Haoyue and Gong, Jinqi and Hong, Richang and Zhang, Min},
  booktitle={Proceedings of the 30th ACM SIGKDD Conference on Knowledge Discovery and Data Mining},
  pages={1062--1072},
  year={2024}
}

@inproceedings{chua2024unified,
  title={Unified Denoising Training for Recommendation},
  author={Chua, Haoyan and Du, Yingpeng and Sun, Zhu and Wang, Ziyan and Zhang, Jie and Ong, Yew-Soon},
  booktitle={Proceedings of the 18th ACM Conference on Recommender Systems},
  pages={612--621},
  year={2024}
}

@inproceedings{gao2022self,
  title={Self-guided learning to denoise for robust recommendation},
  author={Gao, Yunjun and Du, Yuntao and Hu, Yujia and Chen, Lu and Zhu, Xinjun and Fang, Ziquan and Zheng, Baihua},
  booktitle={Proceedings of the 45th international ACM SIGIR conference on research and development in information retrieval},
  pages={1412--1422},
  year={2022}
}

@inproceedings{zhang2025personalized,
  title={Personalized Denoising Implicit Feedback for Robust Recommender System},
  author={Zhang, Kaike and Cao, Qi and Wu, Yunfan and Sun, Fei and Shen, Huawei and Cheng, Xueqi},
  booktitle={Proceedings of the ACM on Web Conference 2025},
  pages={4470--4481},
  year={2025}
}

@inproceedings{wang2023tutorial,
  title={Tutorial: Data Denoising Metrics in Recommender Systems},
  author={Wang, Pengfei and Li, Chenliang and Zou, Lixin and Feng, Zhichao and Li, Kaiyuan and Li, Xiaochen and Liu, Xialong and Wang, Shangguang},
  booktitle={Proceedings of the 32nd ACM International Conference on Information and Knowledge Management},
  pages={5224--5227},
  year={2023}
}

@article{tan2022partial,
  title={Partial relaxed optimal transport for denoised recommendation},
  author={Tan, Yanchao and Member, Carl Yang and Wei, Xiangyu and Wu, Ziyue and Zheng, Xiaolin},
  journal={arXiv preprint arXiv:2204.08619},
  year={2022}
}

@inproceedings{lee2021bootstrapping,
  title={Bootstrapping user and item representations for one-class collaborative filtering},
  author={Lee, Dongha and Kang, SeongKu and Ju, Hyunjun and Park, Chanyoung and Yu, Hwanjo},
  booktitle={Proceedings of the 44th international ACM SIGIR conference on Research and Development in information retrieval},
  pages={317--326},
  year={2021}
}

@article{chen2020bias,
  title={Bias and debias in recommender system: a survey and future directions (2020)},
  author={Chen, Jiawei and Dong, Hande and Wang, Xiang and Feng, Fuli and Wang, Meng and He, Xiangnan},
  journal={arXiv preprint arXiv:2010.03240},
  year={2020}
}

@article{zhang2023robust,
  title={Robust recommender system: a survey and future directions},
  author={Zhang, Kaike and Cao, Qi and Sun, Fei and Wu, Yunfan and Tao, Shuchang and Shen, Huawei and Cheng, Xueqi},
  journal={arXiv preprint arXiv:2309.02057},
  year={2023}
}

@inproceedings{he2016ups,
  title={Ups and downs: Modeling the visual evolution of fashion trends with one-class collaborative filtering},
  author={He, Ruining and McAuley, Julian},
  booktitle={proceedings of the 25th international conference on world wide web},
  pages={507--517},
  year={2016}
}

@article{harper2015movielens,
  title={The movielens datasets: History and context},
  author={Harper, F Maxwell and Konstan, Joseph A},
  journal={Acm transactions on interactive intelligent systems (tiis)},
  volume={5},
  number={4},
  pages={1--19},
  year={2015},
  publisher={Acm New York, NY, USA}
}

@inproceedings{wang2023efficient,
  title={Efficient bi-level optimization for recommendation denoising},
  author={Wang, Zongwei and Gao, Min and Li, Wentao and Yu, Junliang and Guo, Linxin and Yin, Hongzhi},
  booktitle={Proceedings of the 29th ACM SIGKDD conference on knowledge discovery and data mining},
  pages={2502--2511},
  year={2023}
}

\end{document}